\documentclass{PoS}

\usepackage{lineno}
\usepackage{graphicx}
\usepackage{caption}
\usepackage{subcaption}
\usepackage{amsmath,amsfonts,amssymb}
\usepackage{nicefrac}
\usepackage{graphicx,wrapfig,lipsum}
\usepackage{lineno}
\title{ Probing neutrino emission at GeV energies from compact binary mergers with IceCube }

\ShortTitle{Probing neutrino emission at GeV energies from compact binary mergers with IceCube}

\author{
The IceCube Collaboration\footnote{For collaboration list, see PoS(ICRC2019) 1177.}\\
{\itshape \href{http://icecube.wisc.edu/collaboration/authors/icrc19_icecube}{http://icecube.wisc.edu/collaboration/authors/icrc19\_icecube}}\\
E-mail: \email{gdewasse@icecube.wisc.edu}
}

\abstract{
The advent of Multi-Messenger Astronomy has allowed for new types of source searches within the neutrino community. We present the results of the first search for GeV astrophysical neutrinos emitted from Compact Binary Mergers, i.e. binary black hole or binary neutron star mergers, detected by the LIGO and Virgo interferometers. 
We introduce a new approach that lowers the energy threshold of IceCube from roughly 10 GeV to <1 GeV.  This method uses an innovative event selection of GeV neutrino events in IceCube and searches for a statistically significant increase in the amount of GeV-like events detected around the Compact Binary Merger time. We compare our results with constraints set by high-energy neutrino searches, and describe the complementarity of these low and high-energy searches.\\

\vspace{4mm}
{\bfseries Corresponding authors:}
\speaker{Gwenhael de Wasseige}$^{1}$, Imre Bartos$^{2}$, Krijn de Vries$^{3}$, Erin O'Sullivan$^{4}$\\
{$^{1}$ \itshape Laboratoire APC, Paris-Diderot, France}
{$^{2}$ \itshape Department of Physics, University of Florida, USA}
{$^{3}$ \itshape Vrije Universiteit Brussel, Belgium}
{$^{4}$ \itshape Stockholm University and the Oskar Klein Centre}

}

\FullConference{36th International Cosmic Ray Conference -ICRC2019-\\
		July 24th - August 1st, 2019\\
		Madison, WI, U.S.A.}

\begin{document}
\section{Motivation for low-energy astrophysical neutrino searches}
In the past decade, high-energy neutrino telescopes have joined the global effort of multi-messenger astronomy, which aims to detect and study some of the most energetic phenomena in the universe using charged particles, electromagnetic radiation, neutrinos, and gravitational waves. 
High-energy neutrino telescopes are optimized to detect neutrinos in the TeV-PeV range, where astrophysical neutrinos emerge above the background of atmospheric neutrinos created by the interaction of cosmic rays with air nuclei. At lower energies (GeV-TeV), neutrino experiments have so far focused on the characterization of the atmospheric neutrino flux. The GeV energy domain is therefore still a \textit{terra incognita} in terms of astrophysical neutrino observations. Considering that astrophysical neutrino fluxes typically exhibit a power-law decrease with energy, one can expect that neutrino telescopes sensitive to this energy range could allow us to probe larger neutrino fluxes and possibly identify new astrophysical neutrino sources. 

In these proceedings, we will focus on compact binary mergers detected by the LIGO and Virgo interferometers~\cite{ligo}. Part of the detected population may be associated with the detection of a gamma-ray burst (GRB). The neutrino production is expected to come from different processes. While TeV neutrinos are predicted as a consequence of the internal shock in the prompt emission phase of GRBs~\cite{grb-2}, GeV neutrinos would be produced by collisions of neutrons and protons following their decoupling during the acceleration phase~\cite{Kohta}. Besides offering evidence of hadronic acceleration mechanisms, the detection of GeV neutrinos from GRBs would also constitute a probe of the amount of matter surrounding the astrophysical object, allowing better constraints of the environments, acceleration processes, and progenitors of these phenomena.
Until recently, the Super-Kamiokande (SK) detector, optimized for neutrinos in the MeV - GeV range, was the only neutrino detector able to provide limits on the astrophysical neutrino flux in the GeV regime. It has set, among others, limits from the GW170817 binary neutron star merger~\cite{superkamiokande}.

\section{The IceCube Neutrino Observatory: from TeV to GeV neutrino searches}
IceCube is a cubic-kilometer neutrino detector installed in the ice below the geographic South Pole between depths of 1450 m and 2450 m~\cite{jinst}. It consists of 5160 digital optical modules (DOMs) distributed along 86 strings deployed within the detector volume.
A lower energy infill detector, the DeepCore subarray, includes 8 densely instrumented strings with smaller vertical spacing between its optical modules (7 m versus 17 m in the IceCube strings) and smaller horizontal spacing between its strings (72 m in average versus 125 m in IceCube)~\cite{deepcore}. When a neutrino interacts inside or in the neighborhood of the detector, the subsequent electromagnetic and/or hadronic cascade emits Cherenkov photons that can be detected by one or several DOMs.
\begin{figure}[t!h!]
    \centering
        \begin{subfigure}[l]{0.3\textwidth}
                \includegraphics[width=0.95\textwidth]{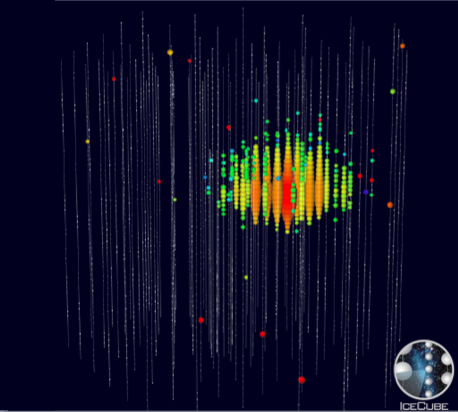}
                \caption{PeV neutrino interaction}
               \label{hese}
        \end{subfigure}
        \begin{subfigure}[c]{0.3\textwidth}
                \includegraphics[width=0.95\textwidth]{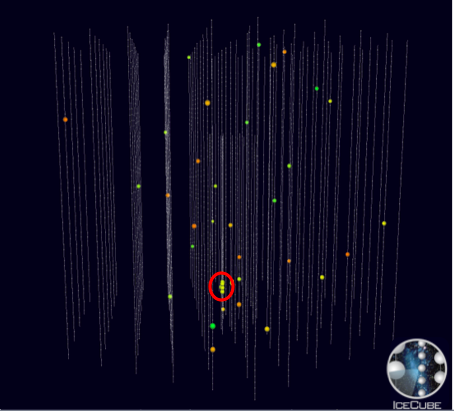}
                \caption{GeV neutrino interaction}
                \label{sf}
        \end{subfigure}
        \begin{subfigure}[r!]{0.3\textwidth}
                \includegraphics[width=0.95\textwidth]{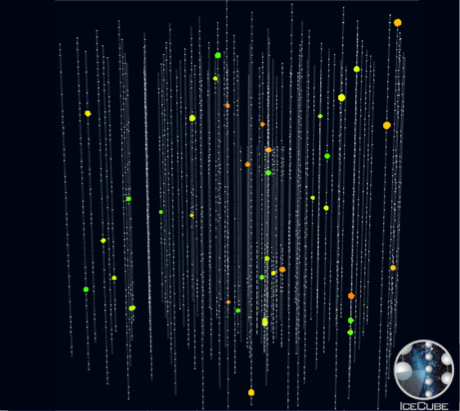}
                \caption{Detector noise event}
                \label{vuvu}
        \end{subfigure}
               \caption{Examples of neutrino interactions in IceCube. A typical GeV neutrino interaction is illustrated within the red circle in \ref{sf} while \ref{hese} and \ref{vuvu}, respectively, show the well-known  "Ernie" event in the PeV range and a detector-noise event (see the text for more details).}\label{events}
\end{figure}
While IceCube was originally dedicated to observe TeV neutrinos, the collaboration has demonstrated the ability to extend the sensitivity to a larger energy range by the use of DeepCore. The collaboration has also joined the worldwide multi-messenger effort studying the highest energetic events in our Universe~\cite{amon, realtime}, leading to the first joint gamma-ray - neutrino observation~\cite{TXS}.

%We present in these proceedings a new approach allowing to extend the sensitivity to neutrinos with an energy around 1~GeV. Using external facilities/experiments to define a time window of interest, as previously described in the case of solar flares, allows IceCube to study astrophysical transient events down to GeV scales.

\subsection{Selection of GeV events}~\label{lowen}
DeepCore allows searches for neutrino interactions down to 10 GeV. Besides a higher density of optical modules, a softer trigger condition has been implemented in DeepCore to improve its sensitivity to lower neutrino energies. We have generated the interactions of GeV neutrinos in IceCube using the interaction physics in GENIE 2.8.6~\cite{genie1}, which includes the nuclear model, cross sections, and hadronization process~\cite{genie2} based on KNO~\cite{genie3} and PYTHIA~\cite{genie4}. The GRV98~\cite{genie5} parton distribution functions are used in the DIS cross sections calculations.

In order to select GeV neutrino interactions, we take advantage of the several existing filters developed to tag specific physics events such as  muons or cascades with an energy larger than 1~TeV ~\cite{jinst}. A GeV interaction shows a very different signature from what is expected from these high-energy events as illustrated in Fig.~\ref{events}. Low-energy events therefore do not pass the high-energy filters. Thus, 1 GeV events can be easily discriminated from the atmospheric muon background in IceCube as well as high-energy neutrino events. Using standard background cuts in IceCube, we reduce the event rate from 1400 Hz to 15 Hz while retaining 98\% of GeV neutrino events.

As one can see in Fig.\ref{events}, the main difference between TeV and GeV neutrino interactions is the amount of light emitted in the ice. 
Putting strong upper constraints on the number of optical modules triggered therefore eliminates neutrinos and remaining muons with an energy exceeding 5~GeV. 
An upper limit on the number of causally connected optical modules, i.e. the DOMs that have likely observed the same physics interaction, further helps to select low energy events. A GeV neutrino interaction produces a small cascade or a short track emitting light close to the interaction vertex, resulting in a small number of causally connected DOMs.
  The main obstacle is identifying GeV neutrino interactions in IceCube among a background due to random hits, which could trigger the detector by mimicking the pattern expected from a low-energy neutrino.
A detailed simulation of noise in the detector helps to estimate the potential contamination of accidental triggers, hereafter referred to as "noise events." These include uncorrelated thermal noise, uncorrelated radioactive noise, and correlated scintillation noise~\cite{larson}.  
About 6~Hz of noise events survives the selection described above, being at this stage the dominant contribution in the event sample. The noise rate can be reduced further by requiring causality between pairs of hits, i.e. requiring that the distance between the two hits divided by the time separation between them is consistent with the speed of light in ice when considering scattering in the ice. Applying the causality condition significantly reduces the rate of detector noise events from 6~Hz to 0.2 Hz.

Additional variables describing the topology of the events help to create a cleaner GeV neutrino sample. Among these, the depth and the charge located around the interaction point as well as the total charge of the event which reduce the data rate down to 0.02~Hz. More than 40\% of the neutrino interactions below 5~GeV generated with GENIE~\cite{genie1} following a generic E$^{-2}$ spectrum pass this selection and are considered in the analysis as GeV event candidates.  

\begin{figure}[t!]
    \centering
  \includegraphics[width=0.85\textwidth]{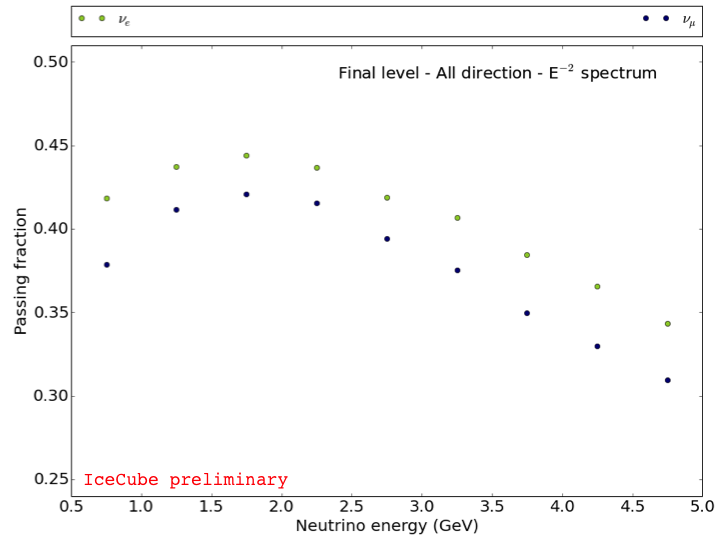}
   \caption{ Passing fraction - ratio of number of events at final level over number of events after triggering - for $\nu_e$ (green) and $\nu_{\mu}$ (blue).\label{passingfraction}}
\end{figure}

\begin{figure}[t!h!]
    \centering
  \includegraphics[width=0.8\textwidth]{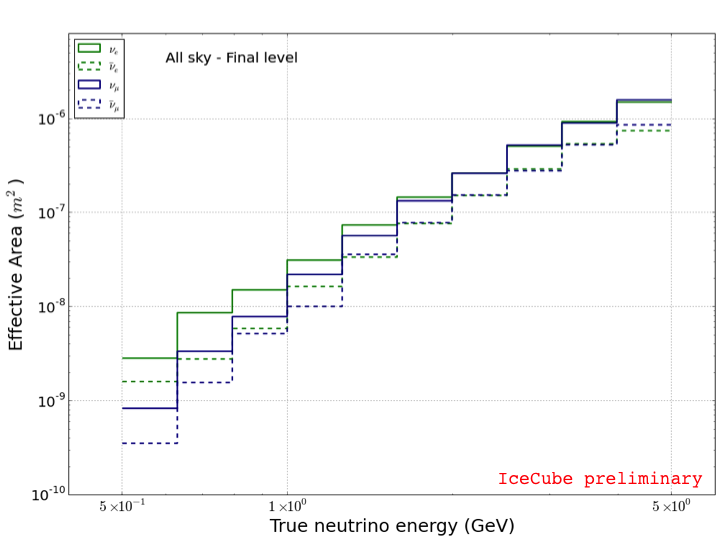}
   \caption{ Effective area for $\nu_e$ (green) and $\nu_{\mu}$ (blue) events at the final level of the event selection. \label{effectivearea}}
\end{figure}

The reduction of the event rate to 0.02 Hz represents a decrease of 5 orders of magnitude from the initial event rate, while keeping more than 40\% of the GeV neutrino events. As shown in Figure~\ref{passingfraction}, the event selection is optimized for neutrinos with an energy between 1 and 2.5~GeV with a slow loss of efficiency at high energies.
The final rate, being dominated by remaining noise events, is slightly larger than the expectation of atmospheric neutrinos, estimated to occur at the mHz level. The IceCube effective area for neutrinos passing the described selection is shown in Figure~\ref{effectivearea}.

\section{Results for selected events}

As previously mentioned, we have used this event selection to search for a GeV neutrino counterpart from compact binary mergers detected by the LIGO and Virgo collaborations (LVC). 
Table~\ref{listofevents} shows the list of events that were analyzed. In the following we will use BBH as abbreviation for binary black hole mergers, BNS for binary neutron star mergers, and NSBH for the merger of a black hole and a neutron star. 

\begin{table}[h!]
    \centering
    \caption{This table shows the compact binary mergers that are considered in this work. We also quote the estimated type of the merger as reported by LVC in the initial GCN, and the (absence of) detection of counterpart.}
    
    \begin{tabular}{c|c|c|c}
    Name & Date - LVC Run & Type & Detected counterpart \\ 
    \hline
        GW150914& 09/14/2015 - O1 & BBH & No \\
        GW151012& 10/12/2015 - O1 & BBH & No \\
        GW151226& 12/26/2015 - O1 & BBH & No \\
        \hline
        GW170104& 01/04/2017 - O2 & BBH & No \\
        GW170608& 06/08/2017 - O2 & BBH & Reported Fermi-LAT subtreshold~\cite{fermi-lat} \\
        GW170814& 08/14/2017 - O2 & BBH & No \\
        GW170917& 08/17/2017 - O2 & BNS & Yes~\cite{grb}\\
        \hline
        S190425z& 04/25/2019 - O3 & BNS & No \\
        S190426c& 04/26/2019 - O3 & NSBH & No \\
        S190510g& 05/10/2019 - O3 & BNS & No \\
            
    \end{tabular}
    \label{listofevents}
    \end{table}
We carried out two different GeV neutrino searches: a search for a prompt signal and a search in an extended time window. The details of the analyses and the corresponding results are described below.

\subsection{Search for a prompt emission}
 This search concerns the LVC events that may be the progenitors of short gamma-ray bursts, so we focus on the BNS and NSBH events. Based on the first BNS detected, GW170817, and the subsequent GRB detected 1.7s later by Fermi-GBM and Swift, we define [t, t + 3s] as a conservative time window for the search of a prompt signal, where t is the merger time reported by LVC.
 
 For each of the analyzed mergers, no events were found in the three seconds 
following the merger time. We therefore set upper limits on the emitted neutrino fluence integrated over 3s, in the 500 MeV - 5 GeV energy range. The upper limit on the fluence at Earth is 1.84 x 10$^7$ neutrinos MeV$^{-1}$ cm$^{-2}$ in the 3 seconds of the search and integrated over the three neutrino flavors.

\begin{figure}[t!]
    \centering
  \includegraphics[width=0.8\textwidth]{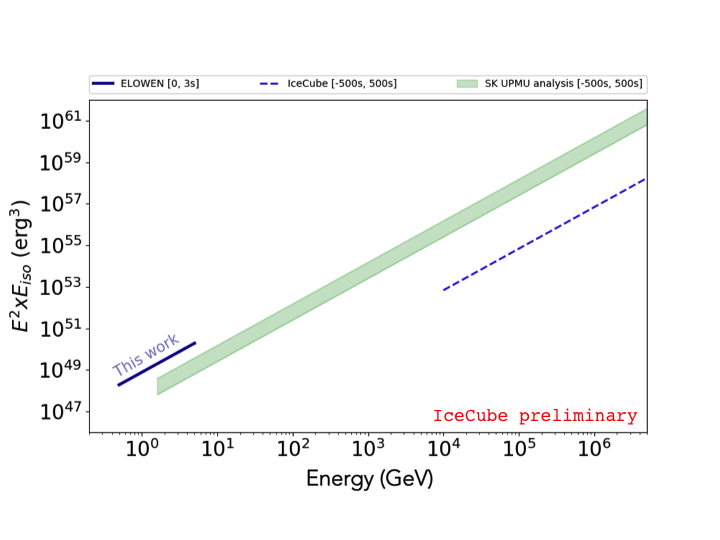}
   \caption{Comparison of $E^2\times E_\mathrm{iso}$ constraints for the neutrino search presented in this work (blue line) with the searches performed by SuperKamiokande (green shaded area) and using high-energy neutrinos (blue shaded area). A time window of 3~s was used for the present work, while 1000~s were integrated to obtain the constraints using high-energy neutrinos and SuperKamiokande data.\label{Eiso2}}
\end{figure}

A convenient way to compare this limit with other constraints obtained for BNS events, is to convert it into a limit on the isotropic-equivalent energy E$_{\mathrm{iso}}$. This variable represents the fluence at Earth multiplied by 4$\pi$ times the squared of the distance between the Earth and the source.
Figure~\ref{Eiso2} shows a comparison of the limit obtained on E$_{\mathrm{iso}}$ with similar constraints obtained by Super-Kamiokande in the 1.6 GeV to 10$^8$ GeV energy range, when focusing on upward-going muons~\cite{superkamiokande}. This analysis found no coincident neutrinos in the time range of $\pm$ 500 s around the time of the gravitational wave event. This null result was used to set an upper limit on the neutrino fluence from the event, which was converted to the upper limit on isotropic energy shown in Figure~\ref{Eiso2}. The ANTARES, Auger and IceCube limits in the 10$^2$ Gev to 10$^{11}$ GeV energy range is also displayed~\cite{icecube-he}. Both analyses have assumed a neutrino production at the source following a power law with a spectral index of -2.
While the limit obtained with the analysis presented in these proceedings is above the other constraints, it probes a lower energy range and thus different production mechanisms as previously described.  We note that the E$_{iso}$ probed by the gamma-ray detection made by Fermi-GBM stands 5 orders of magnitude below the best constraints set using high-energy neutrinos~\cite{icecube-he}. 

\subsection{Search for a counterpart in an extended time window}
\begin{figure}[t!]
    \centering
  \includegraphics[width=0.8\textwidth]{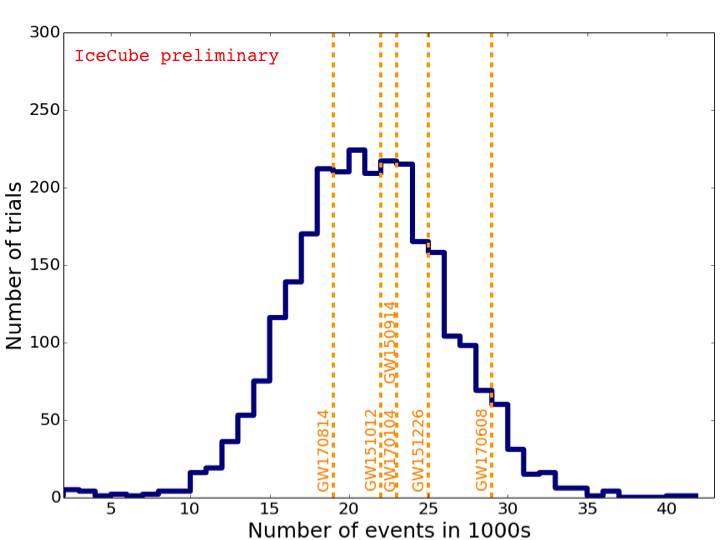}
   \caption{ Background distribution (blue) and BBH O1 and O2 events (orange lines).\label{GW_newunblinding}}
\end{figure}

Using the conservative time window derived in~\cite{gwhen}, we have searched for an excess in the integrated number of events in the [t - 500s, t + 500s] around the merger time t reported by LVC. This search was carried out for some of the binary black hole mergers detected during LVC run O1 and O2. 
Figure~\ref{GW_newunblinding} shows the background distribution obtained using IceCube data when no GRB was reported. The orange lines show the number of events passing our selection for each of the studied BBH mergers. The data recorded during GW170608 is located in the 5\%-tail of the background data distribution, leading to an upper limit on the integrated flux detected at Earth of 50 x 10$^{3}$ neutrinos MeV$^{-1}$ cm$^{-2}$.

%\section{Comparison with existing constraints and observations}

%\begin{figure}[t!h!]
%    \centering
%  \includegraphics[width=0.75\textwidth]{E_iso_comparison.png}
%   \caption{Comparison of E$_{iso}$ constraints for the neutrino search presented in this work with the searches performed by Superkamiokande, the searches for high-energy neutrinos and the observation performed by Fermi-GBM.\label{Eiso}}
%\end{figure}
\section{Summary and Perspectives}
We have presented the first limits obtained in the 0.5 - 5 GeV energy range using the IceCube Neutrino Observatory. While no significant excess events were found for the two searches we performed, we  have validated the innovative approach required to allow IceCube to be sensitive in the GeV energy range and can set low-energy limits on GW and neutrino coincidences which complement existing limits using high-energy neutrinos.
The event selection described in these proceedings will be used to carry out searches for neutrino counterpart in relevant astrophysical phenomena detected via gravitational or electromagnetic waves.

In coming years, we expect the neutrino telescope community to enlarge, with the on-going construction of KM3NeT in the Mediterranean sea~\cite{loi-km3net}, and the deployment of the IceCube-Upgrade~\cite{icrc-upgrade} within IceCube. These detectors will demonstrate a lower detection threshold together with enhanced reconstruction capabilities because of the multi-photomultiplicator geometry of their sensors. It is therefore expected that the limits in the GeV energy range will keep improving. The increasing sensitivity of the gravitational wave interferometers will lead to a larger population to probe using GeV neutrino interactions improving global constraints on the direct environment around the sources.

\end{document}